\documentclass[twocolumn]{svjour3}         % twocolumn
\smartqed  % flush right qed marks, e.g. at end of proof
\usepackage{graphicx}
 \usepackage{mathptmx}      % use Times fonts if available on your TeX system
\begin{document}

\title{A Mid-Infrared Spitzer Study of the Herbig Be Star R Mon and the
Associated HH 39 Herbig-Haro Object%\thanks{Grants or other notes
}

\titlerunning{A Spitzer study of R Mon and HH 39}        % if too long for running head

\author{M. Audard \and
        S. Skinner \and
	M. G\"udel \and
	T. Lanz \and
	F. Paerels \and
	H. Arce
}

\authorrunning{Audard et al.} % if too long for running head

\institute{M. Audard \at
              ISDC \& Geneva Observatory,
	      University of Geneva,
	      Ch. d'Ecogia 16,
	      1290 Versoix,
	      Switzerland \\
              Tel.: +41-22-379-2166,
              Fax: +41-22-379-2133\\
              \email{Marc.Audard@obs.unige.ch}           %  \\
           \and
           S. Skinner \at
              University of Colorado 
	                \and
           M. G\"udel \at
              Paul Scherrer Institut 
	                \and
           T. Lanz \at
              University of Maryland 
	   	                \and
           F. Paerels \at
              Columbia University 
	                \and
           H. Arce \at
              American Museum of Natural History 	      
}

\date{Received: date / Accepted: date}

\maketitle

\begin{abstract}
We report on initial results of our Spitzer Cycle 2 program to observe the young massive star R Mon and its associated HH 39 Herbig-Haro object 
in the mid-infrared. Our program used all instruments on-board Spitzer to obtain deep images with IRAC of the HH 39 complex and of R Mon and 
its surroundings, a deep image of HH 39 at 24 and 70~$\mu$m with MIPS, and mid-infrared spectra with the SH, LH, and LL modules of IRS. The aim of 
this program is to study the physical links in a young massive star between accretion disk, outflows and jets, and shocks in the associated HH 
object. Our preliminary analysis reveals that several knots of HH 39 are clearly detected in most IRAC bands. In IRAC4 (8~$\mu$m), diffuse emission, 
probably from PAHs, appears as foreground emission covering the HH 39 emission. The HH 39 knots are detected at 24 microns,  despite the fact 
that dust continuum emission covers the knots and shows the same structure as observed with IRAC4. The IRS spectra of HH 39 show weak
evidence of [Ne II] 12.8~$\mu$m and 0--0 S(1) H$_2$ 17.0~$\mu$m lines. A more detailed analysis is, however, required due to the faintness of the Herbig-Haro knots. 
Finally, we obtained the SH and MIPS SED spectra of R Mon. A PAH emission feature at 11.3~$\mu$m is detected on top of the strong continuum; 
although no strong emission or absorption lines are observed, we will seek to detect faint lines. The combined IRAC, IRS, and MIPS data of the R 
Mon/HH 39 system will help us to understand circumstellar disk processing, and the connection between jets, outflows, and HH objects. 
\ 
\end{abstract}

\section{Introduction}
\label{intro}

Herbig Ae/Be stars (hereafter HAEBEs; Herbig, 1960) form a class of
massive, young stars with high luminosities ($10-1000~L_\odot$) and
with a strong infrared (IR) excess due to circumstellar dust (Waters and Waelkens, 1998). 
The spectral energy distribution (SED) of HAEBEs can be explained by circumstellar
envelopes with polycyclic aromatic hydrocarbons (PAHs; $3.3$, $6.2$,
$7.7$, and $11.3$~$\mu$m; Brooke et al., 1993; Meeus et al., 2001), and
by dust emission from amorphous or crystalline silicate bands
($8-12$~$\mu$m) or molecular and atomic transitions. Accretion disks
and outflows in the less massive classical T Tau stars (CTTS) are
intimately connected, thus  the presence of outflows in some HAEBEs
 suggests that disks could be present as
well (e.g., Corcoran and Ray, 1998a). Corcoran and Ray (1998b) also found that the wind mass-loss rate
correlates with the IR excess over $5$ orders of magnitude in
luminosity and from $0.5$ to $10~M_\odot$ when using both CTTS and HAEBEs.
Outflows in HAEBEs  are $\sim 2-3$ times faster ($v \sim
600-900$~km~s$^{-1}$) than in CTTS, but generally show similar
collimation ($3^\circ - 10^\circ$), although the fraction of poorly
collimated outflows in HAEBEs ($50^\circ - 120^\circ$) is larger
(e.g., Mundt and Ray, 1994). Complex shocks occur at
the interface between the jet and the molecular material (Draine, 1980),
heating up the gas which in turn cools down radiatively. This gas is
detected as Herbig-Haro objects (HH) in excited lines in the optical,
in the near-IR (e.g., ${\rm H_2}$ ro-vibrational lines around
2~$\mu$m), and in the mid-IR ([O~{\sc i}] 63~$\mu$m; Nisini et al., 1997; Liseau et al., 1997; Molinari et al., 2000).

\section{The R Mon and HH 39 System}
\label{sec:system}

The Herbig Be star R Mon ($d = 800$~pc)
is associated with NGC 2261, a reflection nebula that gradually faints
with increasing wavelength (Close et al., 1997). A bipolar
outflow and a high-velocity jet ($v \sim 100$~km~s$^{-1}$) pointing
toward the nearby HH~39 knots ($r= 7.5^\prime$, ${\rm PA} =
350^\circ$) were detected (Cant\'o et al., 1981; Brugel et al., 1984; Movsessian et al., 2002). 
Evidence for a circumstellar disk is
substantiated as well (e.g., Beckwith et al., 1986; Fuente et al., 2003; Fuente et al., 2006).
A faint CTTS companion at a separation of $0.7^{\prime\prime}$ and with
$K^\prime$ flux of 34~mJy is mentioned by Close et al. (1997) and Weigelt et al.
(2002). The HH~39 cluster of knots covers an area about $25^{\prime\prime} \times 45^{\prime\prime}$ 
(Fig.~\ref{fig:hh39}; Jones and Herbig, 1982; Walsh and Malin, 1985), and the
northernmost knot, knot A, is believed to be the working surface of the jet
(Schwartz and Schultz, 1992).

Near-IR spectra of R Mon exist mostly in the $0.9-2.5$~$\mu$m range
with low ($\sim 400$) spectral power (Kelly et al., 1994; Reipurth and Aspin,
1997; Porter et al., 1998). {\it ISO} neither
observed R Mon nor HH~39. The {\it IRAS}  LRS spectrum ($9-21$~$\mu$m)
of R Mon is classified as an ``unusual spectrum showing a flat 
continuum with unusual features'' (Chen et al., 2000). 
Aspin et al. (1988) and Yamashita et al. (1989) claimed to detect an extended emission eastward of R Mon, but it remained
undetected by Close et al. (1997) with IRTF at 3.16~$\mu$m  
down to $1 \sigma \sim 0.05$~mJy~arcsec$^{-2}$  (CVF filter with an exposure of
60~s). 

\section{Spitzer observations}
\label{sec:obs}

We have obtained 4.2~hrs of {\it Spitzer}  time to study the R Mon/HH~39 system.
Our program (PID 20034) uses {\em all} instruments: IRAC and MIPS images of
HH~39; IRAC images of R Mon and its jets; IRS spectra of R Mon, its jets, and
several of the HH~39 knots; 
and MIPS SED of R Mon. The observations were taken at different epochs, spanning from 
October 2005 to April 2006.

\subsection{The Herbig-Haro object HH 39}
\label{sec:hh39}

\begin{figure}
  \includegraphics[width=0.18\textwidth]{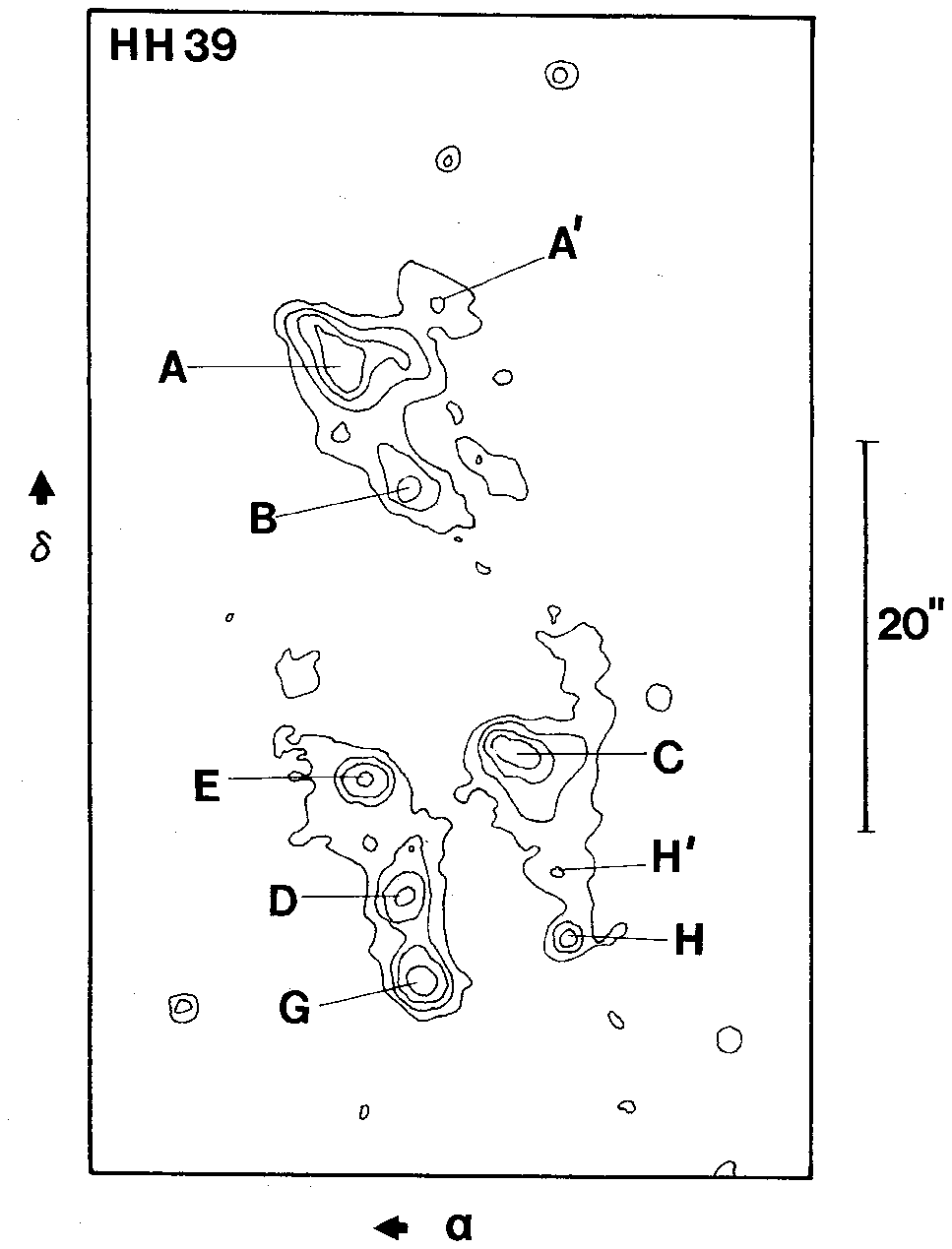}
  \hfill
  \includegraphics[width=0.25\textwidth]{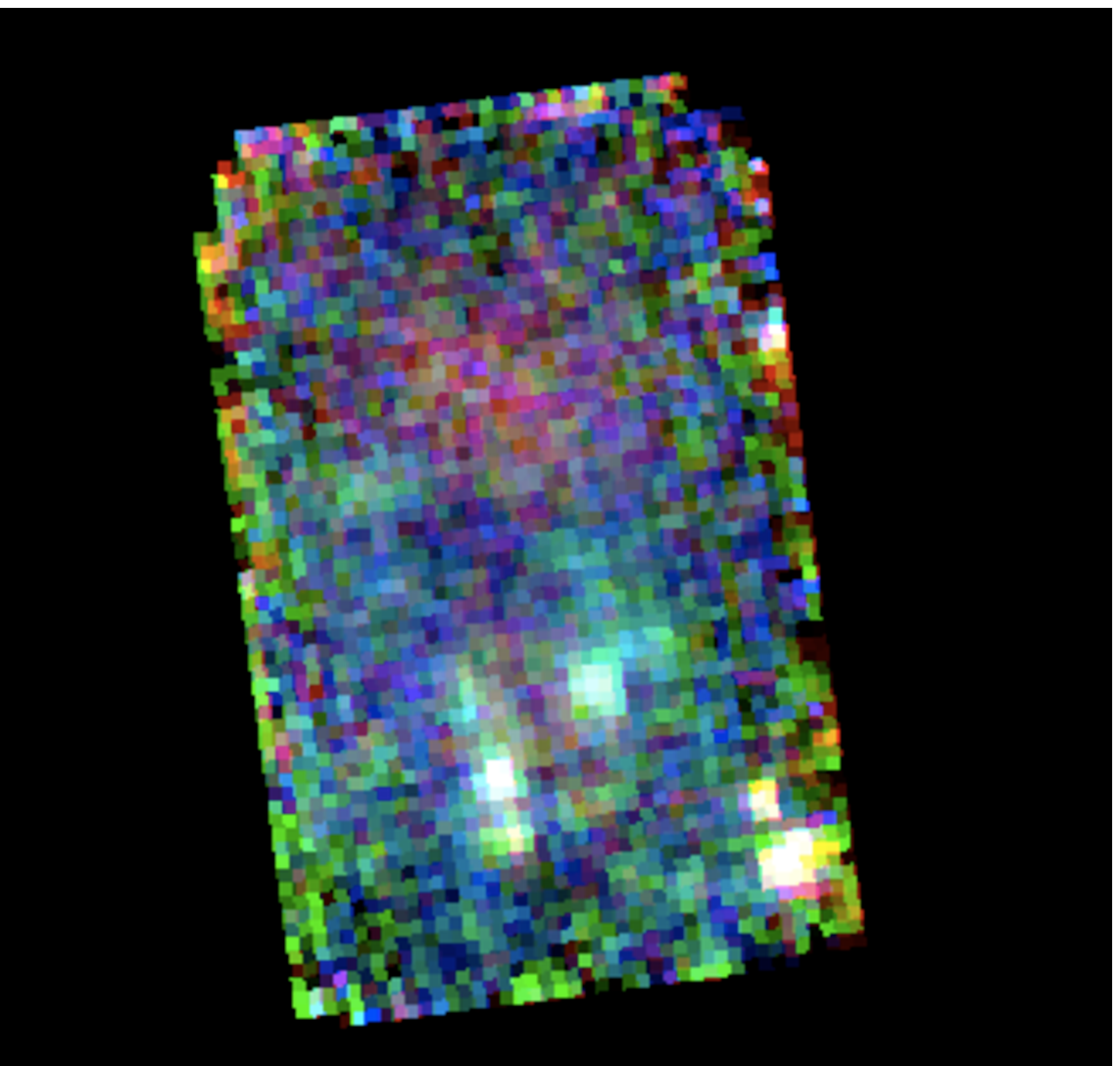}
\caption{(Left) The HH 39 Herbig-Haro knots in the R band  (from Walsh and Malin, 1985).  (Right)
3-color IRAC image of HH 39 (red = 3.6~$\mu$m, 
green = 4.5~$\mu$m, blue = 5.8~$\mu$m).}
\label{fig:hh39}       % Give a unique label
\end{figure}

\begin{figure}
  \includegraphics[width=\linewidth]{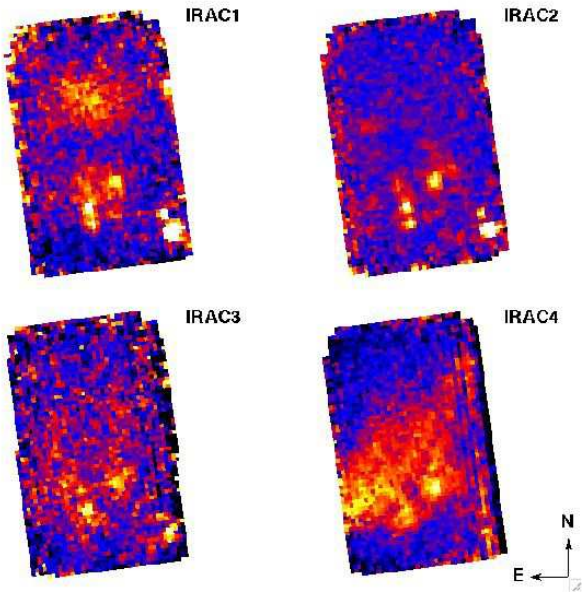}
\caption{(a) IRAC false-color images of HH 39. Background surface flux densities are of the order $0.30, 0.35, 2.1,  
10.5$~MJy~sr$^{-1}$ in IRAC 1, 2, 3, and 4, respectively (zodi estimates: $0.12, 0.37, 
1.72, 10.4$~MJy~sr$^{-1}$; ISM light estimates: $0.15, 0.20, 0.56, 4.79$~MJy~sr$^{-1}$). 
In contrast, the bright knots have peak surface flux densities of about
$0.95, 1.1, 3.2, 11.8$~MJy~sr$^{-1}$. 
}
\label{fig:hh39_irac}       % Give a unique label
\end{figure}

\begin{figure}
  \includegraphics[width=\linewidth]{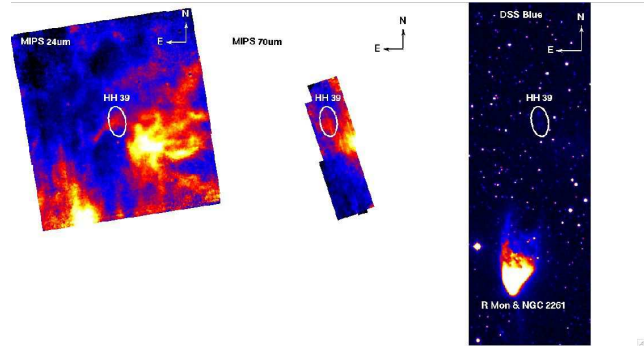}
\caption{MIPS 24 and 70~$\mu$m images (left and middle) of the HH 39 region together with a DSS image (right) of R Mon, 
the reflection nebula NGC 2261 and HH 39, on the same scale.}
\label{fig:hh39_mips}       % Give a unique label
\end{figure}

\begin{figure}
  \includegraphics[width=\linewidth]{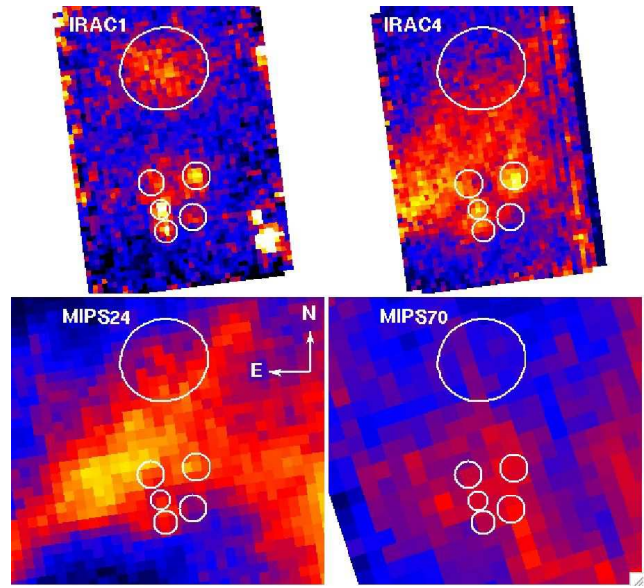}
\caption{IRAC 3.6~$\mu$m and 8~$\mu$m, and MIPS 24~$\mu$m and 70~$\mu$m false-color images together with the positions of HH 39
knots detected in IRAC1. The IRAC 8~$\mu$m and MIPS 24~$\mu$m detect PAH and dust emission across HH 39, likely due to the
upper wall of the NGC 2261 reflection nebula cavity. In contrast, at 70~$\mu$m, the upper wall is not detected, and the HH
39 knots are not clearly detected (see text).}
\label{fig:hh39_irac_mips}       % Give a unique label
\end{figure}

Fig.~\ref{fig:hh39} (left) shows the contour profiles of the Herbig-Haro object HH 39 see in the R band (from Walsh and Malin,
1985). Several knots are labeled, and in particular knot A has been identified as the working surface of the jet onto the
surrounding molecular material. The right panel of fig.~\ref{fig:hh39} shows a 3-color IRAC image of HH~39  (red = 3.6~$\mu$m, 
green = 4.5~$\mu$m, blue = 5.8~$\mu$m). Several knots are clearly detected in the mid-infrared. Fig.~\ref{fig:hh39_irac}
shows false-color images of the HH 39 knots in each of the four IRAC bands. Knots D, G, C, and E are clearly 
detected in all IRAC bands, and knot H is tentatively detected. 
A diffuse emission is observed near the position 
of knot A in IRAC1 only. The feature could be due to
H$_2$ lines or 3.3~$\mu$m PAH. Since this feature is not detected in IRAC2, this is a ``PAH"-free 
band, it is very likely that PAHs contribute to the diffuse emission near knot A. 
H$_2$ lines, on the other hand, most likely contribute to the emission
in knots D, G, C, and E. Supporting evidence comes from a faint 0--0 S(1) rotational H$_2$ line 
detected in the IRS spectra at 17.03~$\mu$m of the bright knots ($0.7-1.0 \times 10^{-20}$~W~cm$^{-2}$). 
In contrast, IRS spectra of knots A+A$^\prime$ do not show evidence 
of H$_2$ line emission in excess of the nearby continuum. Note that faint [Ne II] emission at 
12.81~$\mu$m is measured in the bright knots and at the position of knots A+A$^\prime$ 
(of the order of $0.5 \times 10^{-20}$~W~cm$^{-2}$). At 8~$\mu$m, the bright knots are barely 
visible on top of a diffuse emission that sweeps in through the knots. 
This extended emission is likely of due to PAHs (7.7~$\mu$m) and is also seen in the dust 
continuum 24~$\mu$m MIPS image (Fig.~\ref{fig:hh39_mips} and \ref{fig:hh39_irac_mips}). Its origin is 
unclear but it could be the upper wall of the NGC 2261 reflection nebula cavity.

Fig.~\ref{fig:hh39_mips} shows the MIPS images of the HH~39 region at 24 and 70~$\mu$m together with an optical
Deep Sky Survey image. Dust continuum emission is detected across
the HH 39 knots, in particular at 24~$\mu$m. Nevertheless faint HH 39 knots are also detected at 24~$\mu$m (Fig.~\ref{fig:hh39_irac_mips}), 
but there is no clear evidence of the knots in the 70~$\mu$m image. On the other hand, it should be reminded that the FWHM PRF at 
70~$\mu$m is 18$^{\prime\prime}$, i.e., about the distance from knot A to the group of knots (G, D, E, C, H). Therefore, since the 
MIPS 70~$\mu$m image does not, apparently, observe the same diffuse emission as at 24~$\mu$m or 8~$\mu$m, it is possible that the 
faint emission at the position of the cluster of knots is in fact partially due to [O I] 63~$\mu$m line emission from the knots.

Emission from the upper part of NGC 2261 is also detected in
MIPS at 24~$\mu$m. The bright emission SW of HH~39 is the IR source IRAS 06362+0853. There is no evidence
for diffuse emission in 2MASS around IRAS 06362+0853, suggesting that it is of interstellar cirrus
nature. Surface flux densities at 24~$\mu$m are around 42~MJy~sr$^{-1}$ for the ``blue" color regions and go up to
43.1~MJy~sr$^{-1}$ across HH~39. Peak surface flux densities for IRAS 06362+0853 are about 44.4~MJy~sr$^{-1}$ and
46.7~MJy~sr$^{-1}$ for the dust emission of NGC 2261. At 70~$\mu$m, ``blue" regions are about $31-33$~MJy~sr$^{-1}$,
whereas the emission across HH~39 is $48-50$~MJy~sr$^{-1}$, and $80-85$~MJy~sr$^{-1}$ near IRAS 06362+0853. The
SPOT background estimates are 37.6~MJy~sr$^{-1}$ (zodi) and 2.96~MJy~sr$^{-1}$ (ISM) at 24~$\mu$m and 10.8~MJy~sr$^{-1}$
(zodi), 21.9~MJy~sr$^{-1}$ (ISM), and 0.2~MJy~sr$^{-1}$ (cosmic) at 70~$\mu$m. The surface flux densities in 
the ``blue" regions are of similar values as the estimated total.


\subsection{R Mon}
\label{sec:rmon}

\begin{figure}
  \includegraphics[width=\linewidth]{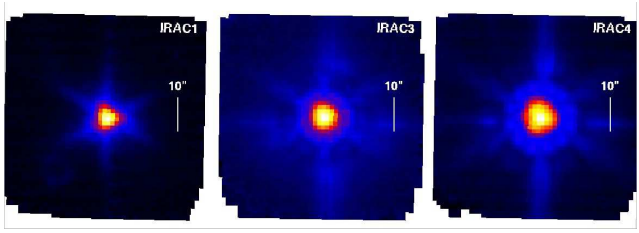}
\caption{IRAC 3.6~$\mu$m, 5.8~$\mu$m, and 8~$\mu$m images (log scale) of the Herbig Be star R Mon. IRAC2 could not be obtained, 
due to R Mon's  brightness at 4.5~$\mu$m, even with a sub-array mode.}
\label{fig:rmon_irac}       % Give a unique label
\end{figure}

Fig.~\ref{fig:rmon_irac} shows the IRAC images of the Herbig Be star R Mon and its immediate surroundings. Note that IRAC2 could not 
be used, even in sub-array mode, due to R Mon's brightness at 4.5~$\mu$m. The main goal of these
observations was to detect any faint emission from the jets (NS direction) or from the circumstellar disk (EW direction). In particular,
we aimed to determine whether a faint eastward extended emission feature detected by Aspin et al.~(1988) and Yamashita et al.~(1989),
but undetected by Close et al.~(1997), could indeed be detected with the highly sensitive IRAC detectors.

The total (zodi and cirrus) background flux densities are $0.35$, $2.86$, $15.5$~MJy~sr$^{-1}$
(based on SPOT). However, the brightness of R Mon is such that the PRF illuminates the full sub-array detector. Aperture photometry
(with a radius of 10 pixels = $12^{\prime\prime}$, requiring no aperture correction), the total flux densities for R Mon are $17.59 \pm 0.29$~Jy, $30.41
\pm 0.40$~Jy, and $31.37 \pm 0.36$~Jy in IRAC 1, 3, and 4, respectively. The estimated background fluxes over a circle of 10 pixel radius
are negligible ($0.004$, $0.03$, $0.17$~Jy) compared to R Mon's brightness and the RMS uncertainties. Note that the above fluxes
include any contribution from R Mon B, separated from R Mon by 0.7$^{\prime\prime}$ (Close et al., 1997). However, in comparison with R Mon, the companion is expected to
contribute negligibly ($0.0013, 0.0085, 0.035$~Jy in JHK$^{\prime}$; Close et al., 1997). The next step will be to subtract the PRF of 
R Mon to possibly detect emission from the disk and the jets. We will need to create IRAC sub-array PRFs from
sub-array data of stars observed, e.g., in the FEPS program.

\begin{figure}
  \includegraphics[width=\linewidth]{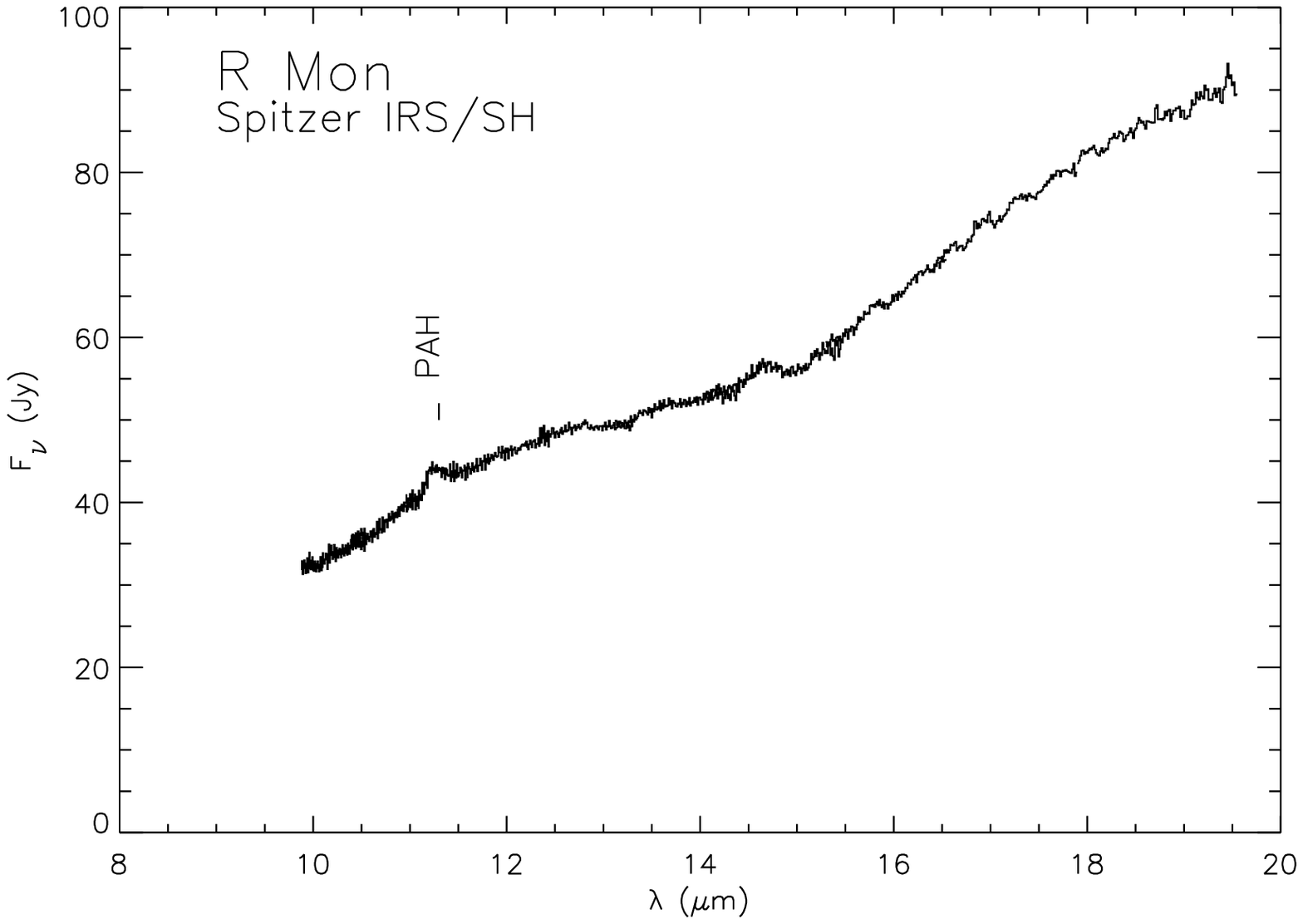}
  \includegraphics[width=\linewidth]{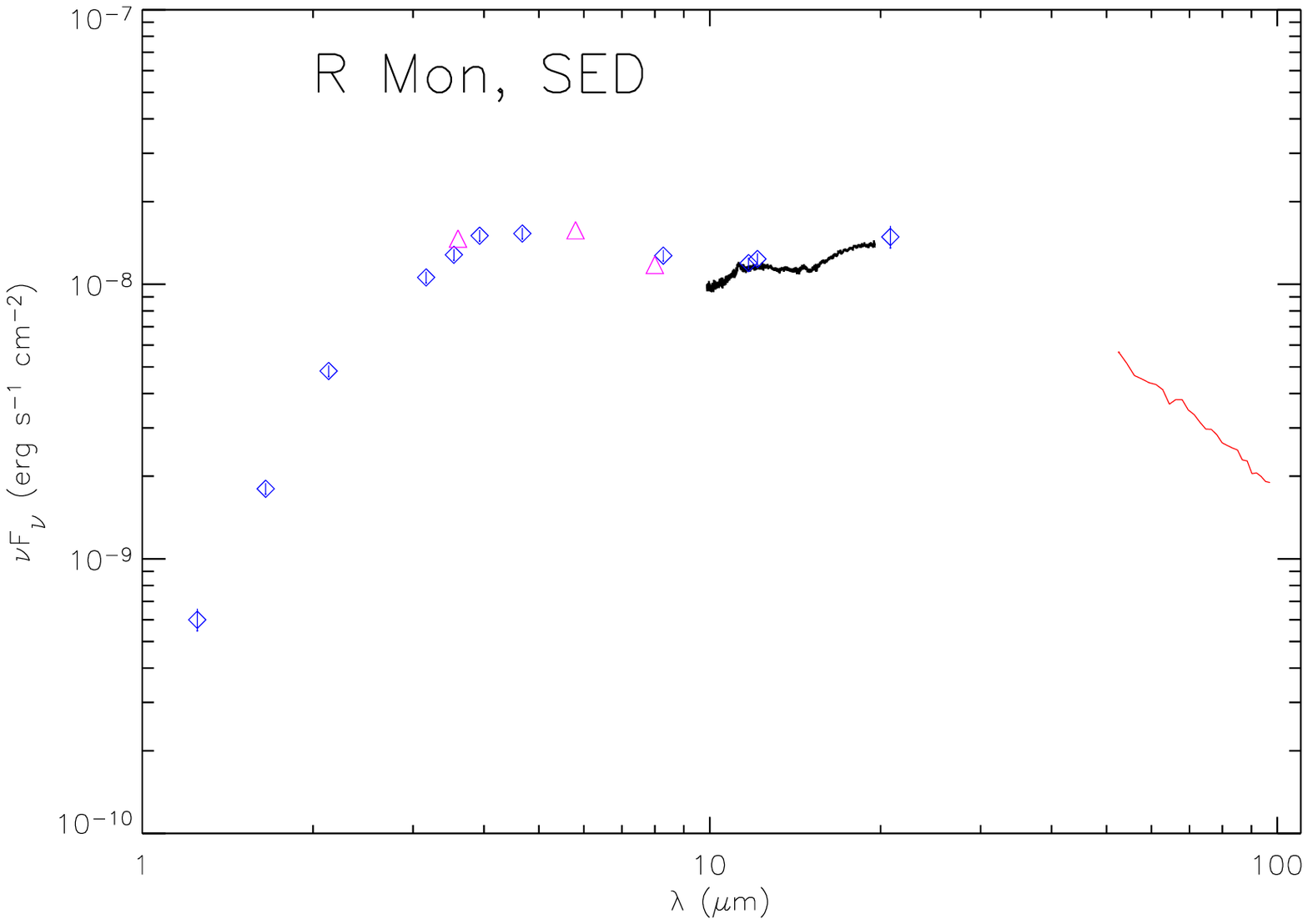}
\caption{(Top): IRS SH spectrum of R Mon. (Bottom): R Mon's raw (not dereddened)
SED from 1 to 100~$\mu$m. IRAC fluxes are shown as triangles (the error bars are
smaller than the symbol), the IRS SH spectrum as a black line, and the MIPS
SED as a red line. Values from the literature (Close et al., 1997 and MSX) are
shown as blue diamonds.}
\label{fig:rmon_spec}       % Give a unique label
\end{figure}

Finally, Fig.~\ref{fig:rmon_spec} shows the IRS SH spectrum of R Mon (top) and its raw SED
from 1 to 100~$\mu$m (bottom). For the latter, we used IRAC fluxes, the IRS SH spectrum, and the MIPS SED and complemented them with values
from the literature (Close et al., 1997; and MSX). The IRS SH spectrum shows a clear PAH feature at $11.3~\mu$m and other faint features
might be present as well. A detailed analysis is ongoing to remove instrumental effects (e.g., defringing).

\begin{acknowledgements}
This work is based on observations made with the Spitzer Space Telescope, which is operated by the Jet Propulsion Laboratory, 
California Institute of Technology under a contract with NASA. Support for this work was provided by NASA through an award (JPLCIT 1275416) 
issued by JPL/Caltech. MA also acknowledges support from a Swiss National
Science Foundation Professorship (PP002--110504). 
\end{acknowledgements}

%\bibliography{biblio}   % name your BibTeX data base

\begin{thebibliography}{}
\bibitem{aspin88}
Aspin, C., et al.: A\&A {\bf 197}, 242 (1988)
\bibitem{beckwith86}
 Beckwith, S., et al.: ApJ {\bf 309}, 755 (1986) 
\bibitem{brooke93}
 Brooke, T.~Y., Tokunaga, A.~T., \& Strom, S.~E.: AJ {\bf 106}, 656 (1993)
\bibitem{brugel84}
 Brugel, E.~W., Mundt, R., \& Buehrke, T.: ApJ {\bf 287}, L73 (1984)
\bibitem{canto81}
 Cant\'o, J., et al.: ApJ {\bf 244}, 102 (1981)
 \bibitem{chen00} 
 Chen, P.~S., Wang, X.~H., \& He, J.~H.: Ap\&SS {\bf 271}, 259 (2000)
\bibitem{close97}
Close, L.~M.~et al.: ApJ {\bf 489}, 210 (1997)
\bibitem{corcoran98a}
 Corcoran, M.~\& Ray, T.~P.: A\&A {\bf 336}, 535 (1998a)
\bibitem{corcoran98b}
 Corcoran, M.~\& Ray, T.~P.: A\&A {\bf 331}, 147 (1998b)
\bibitem{draine80}
 Draine, B.~T.: ApJ, {\bf 241}, 1021 (1980)
\bibitem{fuente03}
 Fuente, A., et al.: ApJ {\bf 598}, L39 (2003)
\bibitem{fuente06} 
Fuente, A., Alonso-Albi, T., Bachiller, R., Natta, A., Testi, L., Neri, R., \& Planesas, P.: ApJ, {\bf 649}, L119 (2006) 
\bibitem{herbig60}
 Herbig, G.~H.: ApJS {\bf 4}, 337 (1960)
\bibitem{jones82}
 Jones, B.~F.~\& Herbig, G.~H.: AJ {\bf 87}, 1223 (1982)
\bibitem{kelly94}
 Kelly, D.~M., Rieke, G.~H., \& Campbell, B.: ApJ {\bf 425}, 231 (1994)
\bibitem{liseau97}
 Liseau, R., et al.: in IAU Symp.~182: Herbig-Haro Flows and the Birth of Stars, {\bf 182}, 111 (1997)
\bibitem{meeus01} 
 Meeus, G., et al.: A\&A {\bf 365}, 476 (2001)
\bibitem{molinari00}
 Molinari, S.~et al.: ApJ {\bf 538}, 698 (2000)
\bibitem{movsessian02}
 Movsessian, T.~A., Magakian, T.~Y., \& Afanasiev, V.~L.: A\&A {\bf 390}, L5 (2002)
\bibitem{mundt94}
 Mundt, R.~\& Ray, T.~P.: in ASP Conf.~Ser.~ 62: The Nature and Evolutionary Status of Herbig Ae/Be Stars, 237 (1994)
\bibitem{nisini96}
 Nisini, B.~et al.: A\&A {\bf 315}, L321 (1996)
\bibitem{porter98}
 Porter, J.~M., Drew, J.~E., \& Lumsden, S.~L.: A\&A {\bf 332}, 999 (1998)
\bibitem{reipurth97}
 Reipurth, B.~\& Aspin, C.: AJ {\bf 114}, 2700 (1997)
\bibitem{schwartz92}
 Schwartz, R.~D.~\& Schultz, A.~S.~B.: AJ {\bf 104}, 220 (1992)
\bibitem{walsh85}
Walsh, J.~R.~\& Malin, D.~F.: MNRAS {\bf 217}, 31 (1985)
\bibitem{waters98}
 Waters, L.~B.~F.~M.~\& Waelkens, C.: ARA\&A {\bf 36}, 233 (1998)
\bibitem{weigelt02}
 Weigelt, G., et al.: A\&A {\bf 392}, 937 (2002)
\bibitem{yamashita89}
Yamashita, T., et al.: ApJ {\bf 336}, 832 (1989)



\end{thebibliography}

%% Non-BibTeX users please use

\end{document}